\documentclass[10pt,conference]{IEEEtran}

\IEEEoverridecommandlockouts

\usepackage{cite}
\usepackage{amsmath,amssymb,amsfonts}
\usepackage{algorithmic}
\usepackage{graphicx}
\usepackage{textcomp}
\usepackage{xcolor}
\usepackage{tabularx}
\def\BibTeX{{\rm B\kern-.05em{\sc i\kern-.025em b}\kern-.08em
    T\kern-.1667em\lower.7ex\hbox{E}\kern-.125emX}}
\usepackage[compatibility=false]{caption}

\usepackage{graphicx}
\usepackage{booktabs} 
\usepackage{subcaption} 
\usepackage{amssymb}  
\usepackage[linesnumbered,ruled,vlined]{algorithm2e}
\usepackage{xspace}
\usepackage{float}
\usepackage{amsthm}

\renewcommand{\footnoterule}{
  \kern-3pt
  \hrule width 0.4\columnwidth height 0.4pt
  \kern 2.6pt
}

\newcommand{\CCSS}{\textit{CCSS}\xspace}
\newcommand{\LIA}{QF\_LIA\xspace}
\newcommand{\HighDiv}{\ensuremath{\mathit{HighDiv}}\xspace}
\newcommand{\SMTLIA}{SMT(LIA)\xspace}
\newcommand{\BAM}{\textit{boundary-aware move}\xspace}

\newtheorem{example}{Example}

\theoremstyle{definition}
\newtheorem{definition}{Definition}

\title{SMT(LIA) Sampling with High Diversity}

\author{
\IEEEauthorblockN{Yong Lai\IEEEauthorrefmark{1}\IEEEauthorrefmark{2}, Junjie Li\IEEEauthorrefmark{1}\IEEEauthorrefmark{2}, and Chuan Luo\IEEEauthorrefmark{3}
}
\IEEEauthorblockA{\IEEEauthorrefmark{1}\textit{College of Computer Science and Technology, Jilin University}, Changchun, China}
\IEEEauthorblockA{\IEEEauthorrefmark{2}\textit{Key Laboratory of Symbolic Computation and Knowledge Engineering Ministry of Education, Jilin
University}, Changchun, China}
\IEEEauthorblockA{\IEEEauthorrefmark{3}\textit{School of Software, Beihang University}, Beijing, China}
}

\begin{document}
\maketitle

\begin{abstract}
SMT sampling refers to the task of generating a set of satisfying assignments (samples) for a given SMT formula. In software testing, SMT sampling techniques have seen increasingly widespread application. An effective SMT sampler should be capable of producing samples with high diversity to maximize coverage of the solution space. However, most SMT samplers struggle to adequately cover the solution space and fail to generate sufficiently diverse solutions.

To address these limitations, we propose \HighDiv, the first iterative sampling framework that integrates CDCL(T) and local search in a bidirectional guided manner. During the bidirectional guidance process, solutions generated by CDCL(T) guide the variable initialization of the local search. Conversely, solutions produced by the local search guide CDCL(T) to further explore the solution space. Additionally, we design a novel local search algorithm, \textit{Context-Constrained Stochastic Search} (\CCSS), which introduces an isolation-based variable initialization strategy and the \BAM operator. These components effectively balance exploration and feasibility throughout the search process.
We conduct an extensive evaluation on QF\_LIA formulas from the SMT-LIB benchmark. The results demonstrate that \HighDiv achieves substantial improvements in diversity over the state-of-the-art SMT sampling tools.
\end{abstract}
\section{Introduction} \label{sec:intro}
Satisfiability Modulo Theories (SMT) is the problem of deciding the satisfiability of a first-order logic formula with respect to certain background theories. Modern software testing techniques extensively utilize SMT solvers to generate assignments that satisfy specific constraints as part of test suite construction~\cite{peleska2011automated,cadar2008klee,DBLP:conf/uss/PoeplauF20}. A typical example is symbolic execution~\cite{cadar2008klee,DBLP:conf/uss/PoeplauF20,godefroid2008automated,godefroid2005dart}, whose core lies in employing SMT solvers to solve path constraints, thereby producing new inputs capable of covering previously unexplored paths. Specifically, symbolic execution leverages SMT solvers to find inputs that satisfy the prefix conditions of program paths, enabling effective coverage of program execution paths.

Traditional symbolic execution techniques typically employ SMT solvers to generate a single satisfying assignment for the constraints of each path prefix. However, generating multiple inputs for the same path prefix may explore different subsequent paths, thereby enhancing test coverage. Studies have shown that obtaining multiple solutions for constraints during symbolic execution can significantly enhance testing effectiveness~\cite{jiang2023evaluating,liu2020legion,huang2020pangolin}.
The task of generating a set of highly diverse solutions for a given SMT formula is known as the \textit{sampling} problem, which has attracted more attention in the community in recent years~\cite{dutra2018smtsampler,dutra2019guidedsampler,peled2023smt,jfs-sampler:icst25}. Meanwhile, SMT sampling techniques have garnered growing interest within the field of software testing~\cite{huang2020pangolin,liu2020legion,DBLP:conf/qrs/RobertGWS21,DBLP:journals/ese/HeradioFGBB22}.

Considering that most programs utilize integer variables and perform linear arithmetic operations, Satisfiability Modulo Linear Integer Arithmetic (\SMTLIA) holds pivotal significance in the fields of software testing and verification~\cite{mccarthy1993towards}. 
For the \SMTLIA sampling problem, although using a solver to enumerate solutions (by adding blocking constraints) for a formula is operationally straightforward, this approach is generally costly, and after multiple invocations, most solvers tend to return similar solutions. This prevalence of similar solutions is largely attributed to the fact that many techniques in solvers (e.g., the general simplex method for linear theories) have a preference for boundary values~\cite{de2008z3,barbosa2022cvc5,cimatti2013mathsat5}. Therefore, there is a practical need to efficiently generate a highly diverse set of solutions for \SMTLIA formulas.

Existing SMT samplers typically treat SMT solvers as black boxes~\cite{peled2023smt,dutra2018smtsampler,dutra2019guidedsampler}. Their core strategy involves obtaining solutions from SMT solvers as initial seeds, which are then used to generate additional solutions. However, this approach inherently limits the diversity of generated samples as a large number of samples originates from a very limited set of seeds.

Recent studies on Boolean satisfiability (SAT) sampling have demonstrated that treating the solving procedure as a white-box and deliberately introducing diversification during the search can substantially improve both effectiveness and efficiency~\cite{luo2021ls,luo2024solving}.
In the context of finding a single solution of SMT problems, Conflict-Driven Clause Learning with Theory (CDCL(T)) and local search are two widely recognized and effective algorithmic frameworks. These two approaches are highly complementary, as demonstrated by their combined use in various recent studies~\cite{cai2022local,zhang2024deep}. Specifically, CDCL(T) leverages its powerful CDCL engine to perform efficient reasoning at the propositional logic level, while local search achieves rapid solving by flexibly modifying the current assignment.
However, existing SMT solvers only focus on quickly finding a satisfying assignment without consideration of solution diversity during assignment generation. Consequently, striking an appropriate balance between sample diversity and search efficiency remains a challenge in \SMTLIA sampling tasks.
This raises a fundamental question: 
\emph{Can we combine CDCL(T) with local search in a white-box way to design high-diversity \SMTLIA sampling?}

In this paper, we introduce a novel sampling framework, \HighDiv, designed to tackle the challenges inherent in \SMTLIA sampling. The framework utilizes bidirectional guidance between local search and CDCL(T) for iterative sampling, aiming to enhance sample diversity while maintaining solving efficiency. Specifically, the solution from CDCL(T) guides variable initialization during the local search phase, while the solution generated by local search is fed back into CDCL(T) to further explore the solution space. 

In addition, for the local search component, we designed Context-Constrained Stochastic Search (\CCSS), which introduces a novel variable initialization strategy called the \textit{isolation-based variable initialization}. This strategy partitions the constraints into different subsystems and initializes each subsystem based on its specific characteristics. Furthermore, to allow more flexible modifications of integer variables during the search, we introduced a new operator, \BAM. This operator performs a random move while satisfying the corresponding constraints with a certain probability. For the CDCL(T) component, we modified its \textit{branching heuristic} and \textit{phase selection heuristic}, enabling more flexible exploration of the solution space at the propositional level.

Our experiments demonstrate the diversity of samples generated by \HighDiv in terms of coverage on the abstract syntax tree of SMT formulas~\cite{peled2023smt}. 
Within 900 seconds of experimental time, \HighDiv achieves an average coverage of 53.36\%, while the state-of-the-art \SMTLIA samplers, MeGASampler~\cite{peled2023smt} with its two sampling strategies (random sampling and blocking sampling) and SMTSampler(Int)~\cite{dutra2018smtsampler} achieve average coverages of 33.79\%, 34.88\%, and 34.69\%, respectively. 
When generating a sample set of fixed size 1000, \HighDiv reaches an average coverage of 53.43\%, whereas two versions of MeGASampler and SMTSampler(Int) achieve average coverages of 22.98\%, 20.78\%, and 34.25\%, respectively. 
These results clearly show that \HighDiv is of significantly higher diversity than the state-of-the-art \SMTLIA samplers. 

The primary contributions of this work are outlined below.

\begin{itemize}
  \item We propose \HighDiv, a \SMTLIA sampling framework that combines CDCL(T) and local search with bidirectional guided iteration, and treats the search process as a white-box, thereby advancing \SMTLIA sampling.
  \item We investigate the trade-off between sample diversity and search efficiency in our local search algorithm, Context-Constrained Stochastic Search (\CCSS), featuring a novel variable initialization strategy and the \BAM operator.
  \item We implement \HighDiv and compare it against state-of-the-art \SMTLIA samplers. The results demonstrate that \HighDiv generates more diverse samples under both fixed-time and fixed-sample-size conditions.
\end{itemize}

The structure of this paper is as follows: Section~\ref{sec:pre} introduces the relevant definitions and the frameworks of local search and CDCL(T); Section~\ref{sec:framework} presents the overall framework of \HighDiv, as well as the interaction between stochastic CDCL(T) and \CCSS; Section~\ref{sec:cdcl} introduces the stochastic CDCL(T) component; Section~\ref{sec:ls} provides a detailed explanation of \CCSS; Section~\ref{sec:experiment} presents the experiments; Section~\ref{sec:related} discusses related work; Section~\ref{sec:conclusion} concludes.
\section{Preliminary}\label{sec:pre}

In this section, we introduce the theory of Linear Integer Arithmetic (LIA) in SMT and search techniques to decide its satisfiability, including Conflict-Driven Clause Learning with Theory, also known as CDCL(T), and the local search framework for \SMTLIA.

\subsection{\SMTLIA}
A formula under the Linear Integer Arithmetic (LIA) theory consists of a set of \textit{atomic formulas}, where an atomic formula can be a propositional variable or an arithmetic formula. Arithmetic formulas can be consistently expressed in the standard form $\sum_{i=0}^{n-1} a_ix_i \le k$ or $\sum_{i=0}^{n-1} a_ix_i = k$, where $x$ represents an integer variable, with $a$ and $k$ as constants. A \textit{literal} can be either an atomic formula or its negated form. A \textit{clause} is composed of the disjunction of a set of literals, and a formula in \textit{Conjunctive Normal Form} (CNF) consists of the conjunction of a set of such clauses. Given sets of propositional variables $P$ and integer variables $X$, which are integral components of the \SMTLIA formula $F$, an \emph{assignment} $\alpha$ of $F$ maps each $x \in X$ to an integer in $\mathbb{Z}$ and each $p \in P$ to $\mathit{true}$ or $\mathit{false}$. Under this assignment, $\alpha(x)$ and $\alpha(p)$ represent the values of $x$ and $p$, respectively.

The (Boolean) \textit{skeleton} of the SMT formula $F$ is obtained by replacing each atomic formula $\sigma$ in $F$ with its uniquely corresponding Boolean variable $p_\sigma$, where $p_\sigma$ is referred to as the (Boolean) encoder of $\sigma$.

\begin{example}\label{ex:SMT(LIA)}
Given a set of integer variables $X = \{x_1, x_2, x_3, x_4\}$ and a set of Boolean variables $P = \{p_1, p_2\}$, the following $F_{SMT(LIA)}$ is an \SMTLIA CNF formula,
\begin{equation*}
\begin{split}
F_{SMT(LIA)} =  &\quad (p_1 \lor \neg p_2) \\
				&\land \left( \neg (x_1 + x_2 \le 2) \lor (-2x_1 + 3x_3 \leq 0) \right) \\
				&\land \left( p_2 \lor \left(3x_2 - 7x_3 \le 3 \right) \right).
\end{split}
\end{equation*}
We can construct the following Boolean skeleton $S_F$ of $F_{SMT(LIA)}$ by introducing new Boolean variables $p_{\sigma_1}, p_{\sigma_2}, p_{\sigma_3}$, corresponding to each LIA atomic formula.
\begin{equation*}
\begin{split}
	S_F = &\quad (p_1 \lor \neg p_2) \land \left( \neg p_{\sigma_1} \lor p_{\sigma_2} \right) \land (p_2 \lor p_{\sigma_3})
\end{split}
\end{equation*}
\end{example}

\subsection{CDCL(T) Framework}
Most current SMT solvers, such as Z3~\cite{de2008z3} and CVC5~\cite{barbosa2022cvc5}, deal with \SMTLIA formulas mainly using the Conflict-Driven Clause Learning with Theory (CDCL(T)) algorithm. Within the CDCL(T) framework, a SAT solver based on the Conflict-Driven Clause Learning (CDCL) algorithm is used to reason about the Boolean skeleton of the \SMTLIA formula and solve this Boolean skeleton. The assignment generated by the SAT solver is then transferred to the theory solver, which processes the conjunction of the corresponding LIA atoms. The theory solver checks the consistency of the assignment under LIA theory and performs theory-based deductions.

\begin{algorithm}[htbp]
	\caption{CDCL(T) Algorithm.}\label{alg:CDCL(T)}
	\KwIn{%
		\begin{tabular}[t]{@{}l@{}}
			$F$: a SMT formula;
		\end{tabular}
	}
	\KwOut{the solution of $F$, or reporting "UNSAT";}
	$\alpha \leftarrow \{\}$;\\
	\While{true}{
		$c \leftarrow propagate()$;\\
		\uIf{$c \ne \emptyset$}{
			$lvl \leftarrow resolve\_conflict(c)$;\\
			\If{$ lvl < 0$}{
				\Return{'UNSAT'};
			}
			$backtrack(lvl)$;
		}
		\Else{
			\If{!decide()}{
				\Return{$\alpha$};
			}
		}
	}
	\Return $\alpha$;
\end{algorithm}

Algorithm~\ref{alg:CDCL(T)} demonstrates the CDCL(T) framework implemented in Z3~\cite{de2008z3}. When a formula $F$ is loaded, its Boolean skeleton $S_F$ is abstracted. The assignment $\alpha_S$ to $S_F$ is continuously maintained throughout the search process of the CDCL(T) algorithm.

In line 3, \textit{propagate()} reasons about the unassigned variables in $S_F$ based on the current assignment. A new learnt clause is added to the clause database whenever a conflict is detected during propagation (line 3). At this point, the algorithm processes the conflict, derives the backtracking level \textit{lvl}, and clears certain variable assignments associated with the conflict (line 5). If the backtracking level $lvl < 0$ is deduced from the conflict, the algorithm returns 'UNSAT'.

Alternatively, if no further implications can be derived during the propagation process and no conflicts arise under assignment $\alpha_S$, an unassigned variable is selected based on the \textit{branching heuristic} and assigned according to the \textit{phase selection heuristic} (line 10). Once a complete assignment is found without conflicts, the algorithm returns the current assignment $\alpha$, which serves as a \textit{model} of the formula. More details about the CDCL(T) algorithm can be found in the references~\cite{kroening2016decision,barrett2021satisfiability,ganzinger2004dpll}.

\subsection{Local Search Framework for \SMTLIA}

The local search component of \HighDiv uses the two-mode framework of LS-LIA, the first local search algorithm for \SMTLIA~\cite{cai2022local}. 
We briefly introduce the LS-LIA.

\begin{algorithm}[htbp]
	\caption{Integer Mode of LS-LIA.}\label{alg:ls-lia}
	\While{non\_impr\_steps $\le L \times P_i$}{
		\If{all clauses are satisfied}{
			\Return $\alpha$;
		}
		\uIf{$\exists$ decreasing cm operation}{
			$op :=$ such an operation with the greatest $score$;
		}
		\Else{
			update clauses weights;\\
			$c :=$ a random falsified clause with integer variable;\\
			$op :=$ a $cm$ operation in $c$ with greatest $score$;
		}
		$\alpha := \alpha$ with $op$ performed;
	}
\end{algorithm}

This algorithm divides its search process into two modes: \textit{Boolean mode} and \textit{Integer mode}. In different modes, operations on variables of the appropriate data type are selected to modify the current assignment. In each mode, when the number of \textit{non-improving steps} reaches the threshold, it switches to another mode. This threshold is defined as $L \times P_b$ for the Boolean mode and $L \times P_i$ for the Integer mode, where $P_b$ and $P_i$ represent the proportion of Boolean and Integer literals, respectively, in the falsified clauses, and $L$ is a parameter.

Within a local search algorithm, the pivotal element is the \textit{operator}, which prescribes the manner in which the current solution may be altered. Once the operator is concretized by selecting a particular variable and designating a value to assign, it materializes as a specific \textit{operation}. In Boolean mode, the \textit{flip} operator turns a Boolean variable to its opposite value. The Integer mode, as described in Algorithm~\ref{alg:ls-lia}, introduces a unique operator called the \textit{critical move} (cm), which is defined below.

\begin{definition}\label{def:cm}
The critical move operator, represented as $cm(x, \ell)$, sets the integer variable $x$ to a threshold value that satisfies the literal $\ell$, where $\ell$ is a falsified literal that includes $x$.
\end{definition}

The threshold mentioned above refers to the minimal modification required to make the literal $\ell$ true for $x$. Example~\ref{ex:cm} is provided to assist readers in understanding the definition.

\begin{example}\label{ex:cm}
Given the \SMTLIA formula, which includes a literal, $\ell: (x_1 - 5x_2 \le -5)$, the initial variable assignments are $\{x_1 = 0, x_2 = 0\}$. The operations $cm(x_1, \ell)$ and $cm(x_2, \ell)$ involve assigning $-5$ to $x_1$ and $1$ to $x_2$, respectively, making the literal $\ell$ true.
\end{example}

Whenever a $cm(x, \ell)$ operation is performed, the corresponding literal $\ell$ is set to $\mathit{true}$. Therefore, during the algorithm search process, falsified literals are consistently selected and a $cm$ operation is applied to make $\ell$ true.

\begin{definition}\label{def:score}
For an operation $op$, we define $score(op)$ as the reduction in the total penalty weight of falsified clauses after applying $op$.
\end{definition}

An operation is \textit{decreasing} if its score is greater than 0. For further details on LS-LIA, please refer to~\cite{cai2022local}.

\section{The Framework of \HighDiv} \label{sec:framework}

\begin{figure*}[!ht]
    \centering
    \includegraphics[width=\textwidth, trim=0cm 6.5cm 2.5cm 6cm, clip]{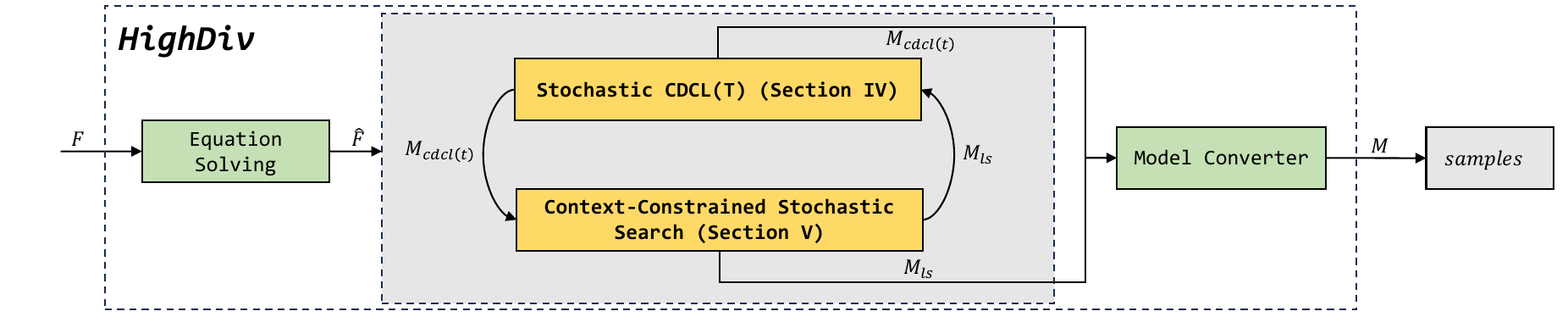}
    \caption{Framework of \HighDiv}
    \label{fig:HighDiv-framework}
\end{figure*}

We combine CDCL(T) and local search in a white-box way to design a high-diversity \SMTLIA sampling method called \HighDiv.
Figure~\ref{fig:HighDiv-framework} provides an overview of the general structure and sequence of operations within \HighDiv. The sampling framework is composed of two distinct components: stochastic CDCL(T) (Section~\ref{sec:cdcl}) and Context-Constrained Stochastic Search (\CCSS), a novel local search algorithm for sampling \SMTLIA formulas and achieving high diversity (Section~\ref{sec:ls}). The two components interact through bidirectional guidance: the solutions generated by \CCSS dynamically guide the stochastic CDCL(T) to further explore the solution space, while the solutions obtained from the stochastic CDCL(T) are used to guide the variable initialization in \CCSS.

Algorithm~\ref{alg:highdiv_framework} presents the top-level framework of \HighDiv. It takes two inputs:1) the \SMTLIA formula \(F\), and 2) the desired number of solutions \(k\).

\HighDiv alternates between a CDCL(T) phase (lines 6-11) and a \CCSS phase (lines 13-17).
First, Z3’s solve-eqs tactic~\cite{de2008z3} applies Gaussian elimination to simplify equalities, producing the pre-processed formula $\hat F$ (line 2). Next, a stochastic CDCL(T) solver attempts to solve the current under-approximation $\hat F_{under}$. If successful, the resulting model $M_{cdcl(t)}$ is mapped back to the original formula $F$ and added to $\mathit{samples}$ (lines 8-9). Guided by $M_{cdcl(t)}$, the algorithm then executes \CCSS on $\hat F$ (line 13). Upon success, \CCSS returns a model $M_{ls}$, which is likewise converted and appended to $\mathit{samples}$ (lines 15-16).
To build the next under-approximation formula, each variable in $\hat F$ is independently selected with probability 0.5 and fixed to its value in $M_{ls}$, yielding $\hat F_{under}$ (line 17). These two phases iterate, enlarging $\mathit{samples}$ until the desired sample count is reached ($|\mathit{samples}|=k$) or the time limit expires.

\begin{algorithm}[htbp]
	\caption{The Framework of \HighDiv}\label{alg:highdiv_framework}
	\KwIn{%
		\begin{tabular}[t]{@{}l@{}}
			$F$: An \SMTLIA formula;\\
                $k$: Allowed number of samples;
		\end{tabular}
	}
	\KwOut{$\mathit{samples}$: The set of samples, i.e., solutions of $F$;}
        $\mathit{samples} \gets \emptyset$; \\
        $\hat{F} \leftarrow equation\_solving(F)$; \\
        $\hat{F}_{under} \gets \hat{F}$; \\
	\While{$|\mathit{samples}| < k$ \textbf{and} time limit not reached}{
            /* Stochastic CDCL(T) */ \\
            $M_{cdcl(t)} \gets stochastic\_cdcl(t)\_solving(\hat{F}_{under})$; \\
            \If{$M_{cdcl(t)} \neq \emptyset$}{
                $M \leftarrow model\_convert(M_{cdcl(t)}, F)$; \\ 
                $\mathit{samples} \gets \mathit{samples} \cup \{M\}$; \\
                \If {$|\mathit{samples}| = k$}{
                    \Return $S$;
                }
            }

            /* Context-Constrained Stochastic Search */ \\
            $M_{ls} \gets CCSS(\hat{F}, M_{cdcl(t)})$; \\
            \If{$M_{ls} \neq \emptyset$}{
                $M \leftarrow model\_convert(M_{ls}, F)$; \\ 
                $\mathit{samples} \gets \mathit{samples} \cup \{M\}$; \\
                $\hat{F}_{under} \gets fix\_partial\_assignment(\hat{F}, M_{ls})$;
            }
	}
	\Return $\mathit{samples}$;
\end{algorithm}
\section{The Stochastic CDCL(T) Component} \label{sec:cdcl}

Previous studies have demonstrated that CDCL(T) and local search methods complement each other effectively~\cite{cai2022local}. A primary reason is the powerful reasoning capabilities of the CDCL engine at the propositional logic level. Inspired by SAT sampling algorithms based on stochastic CDCL~\cite{golia2021designing}, we implemented the stochastic CDCL(T) algorithm in Z3. Specifically, we modified the \textit{branching heuristic} and \textit{phase selection heuristic} in Z3, randomly selecting undecided variables at each decision point and assigning them a random phase (Algorithm~\ref{alg:CDCL(T)}, line 10).

Stochastic CDCL(T) plays a pivotal role in \HighDiv. On one hand, the model generated by CDCL(T) can guide the variable initialization in \CCSS; on the other hand, the integrated theory solvers within CDCL(T), such as the general simplex method, offer a perspective on the solution space that differs from \CCSS, thereby enabling \HighDiv to produce more diverse solutions.
\section{Context-Constrained Stochastic Search} \label{sec:ls}
In this section, we first introduce Context-Constrained Stochastic Search (\CCSS); we then detail its variable-initialization strategy and the \BAM operator, and finally demonstrate how these components jointly mediate the trade-off between sample diversity and algorithmic scalability.

\begin{algorithm}[htbp]
	\caption{Integer Mode of \CCSS}
        \label{alg:sls}
	\KwIn{%
		\begin{tabular}[t]{@{}l@{}}
			$F$: SMT(LIA) formula;\\
			$M$: A model of $F$;\\
		\end{tabular}
	}
	\KwOut{$\alpha$: the solution of $F$;}
	$isolation\_based\_variable\_initialization(\alpha, M)$;\\
	
	\While{$Step_{total} \le $ MaxSteps}{
    	\If{$\alpha$ satisfies $F$}{
    		\Return $\alpha$;
    	}
            $B \leftarrow \emptyset$;\\
            \ForEach{literal $\ell$ in falsified clauses}{
                $B \leftarrow B \cup \{bam(x,\ell,\mathcal{L}^{\text{ctx}}(x)) \mid x \text{ appears in } \ell \};$\\
            }
		  \If{$\exists$ decreasing $bam$ operation in B}{
			$op$ $\leftarrow$ select such an operation with the greatest $score$;
		  }
		  \Else{
			update clauses penalty weights; \\
			$c \leftarrow $ a random falsified clause with integer variable; \\
			$op \leftarrow$ the $bam$ operation with the greatest score in c;\\
		  }
		  $\alpha \leftarrow \alpha$ with $op$ performed;
	}
	\Return $\alpha$;
\end{algorithm}

\CCSS uses the two-mode framework of LS-LIA~\cite{cai2022local}. 
Algorithm~\ref{alg:sls} describes the Integer mode of \CCSS algorithm. Line 1 generates the initial assignment $\alpha$ using the \textit{isolation-based variable initialization strategy}. Lines 2-14 iteratively refine $\alpha$ until the number of consecutive non-improving steps exceeds the threshold $\mathit{MaxSteps}$ or the current assignment satisfies the formula $F$. In lines 6-7, \textit{boundary-aware moves} are constructed for every literal in the falsified clause and inserted into the candidate set $B$. Each iteration selects from $B$ the highest-scoring \BAM that decreases the number of falsified clauses (lines 8-9). If no such decreasing \BAM exists, the search is deemed stuck in a local optimum; the algorithm then updates the penalty weights (line 11) and picks a random falsified clause with integer literals and chooses a \BAM operation with the greatest score (lines 12-13).

In Boolean mode, the formula reduces to a purely propositional form. Consequently, we retain the LS-LIA search strategy and randomly assign Boolean variables solely during the initialization phase.

\subsection{Isolation-based Variable Initialization Strategy}

In local search algorithms, the quality of variable initializations significantly impacts search efficiency~\cite{frohlich2015stochastic,liu2025optimizing}. Traditional variable initialization methods typically rely solely on boundary information provided by the formula~\cite{cai2022local}. In the context of sampling problems, when initializing variables, it is essential to consider not only the impact of the initial values on search efficiency but also the potential detrimental effect of a uniform initialization on the diversity of the solutions. 

When solving constraint satisfaction problems, the constraint system can typically be decomposed into multiple independent subsystems for tailored solving.
We propose a similar strategy called \emph{isolation-based variable initialization} to obtain an initialized assignment in line 1 in the \CCSS algorithm.
Specifically, we partition the integer variables into three independent constraint subsystems: the \textit{equality system}, the \textit{high-frequency system}, and the \textit{general system}.
Then, decomposing \SMTLIA formulas into multiple sub-constraint systems and initializing variables based on the specific properties of the constraints achieves a better balance between sample diversity and search efficiency. 

\subsubsection{Initialization of Variables in Equation Systems}

\begin{algorithm}[htbp]
\caption{Closure of the Equality System Variable Set $\mathcal{E}_{eq}$}\label{alg:eq_sys}
\KwIn{$\mathcal{C}$: set of constraints}
\KwOut{$\mathcal{E}_{eq}$: variables in the equality system}

$\mathcal{C}_{=} \leftarrow equality\_constraints(\mathcal{C})$\;
$\mathcal{C}_{\neq} \leftarrow \mathcal{C} \setminus \mathcal{C}_{=}$\;
$\mathcal{E}_{eq} \leftarrow \displaystyle\bigcup_{c\in \mathcal{C}_{=}} \operatorname{vars}(c)$\;

\Repeat{no variable is added to $\mathcal{E}_{eq}$}{
    \ForEach{$c \in \mathcal{C}_{\neq}$}{
        \If{$\operatorname{vars}(c)\cap\mathcal{E} \neq \varnothing$}{
            $\mathcal{E}_{eq} \leftarrow \mathcal{E} \cup \operatorname{vars}(c)$
        }
    }
}
\Return $\mathcal{E}_{eq}$\;
\end{algorithm}

Equality constraints are more stringent than inequality constraints, thus requiring special handling during the solving process. Since \HighDiv is integrated into Z3, it is convenient to use the equality solving preprocessing tactic provided by Z3, namely \textit{solve-eqs}~\cite{de2008z3}. However, this tactic only performs Gaussian elimination on the equality constraints in the conjunction. As a result, some equality constraints will still remain and be carried over into the subsequent local search phase.

To improve the efficiency of solving equality constraints, we construct a closed equality system within the original constraints. Algorithm~\ref{alg:eq_sys} initializes $\mathcal{E}_{eq}$ with every variable that appears in an equality constraint (line 3).
It then repeatedly scans the non-equality constraints; whenever such a constraint shares at least one variable with the current $\mathcal{E}_{eq}$, all of its variables are added to $\mathcal{E}_{eq}$ (lines 4-8).
The process stops when a full scan adds nothing new, yielding the transitive closure of variables connected to any equality constraint.

Clearly, this equality system is independent, meaning that the behavior of variables outside the equality system will not affect the equality system during subsequent computations. Therefore, when initializing variables in the equality system, \HighDiv reduces the computational overhead in the subsequent solving process by setting all variables to zero.

\subsubsection{Initialization of Variables in High-frequency System}

\begin{algorithm}[htbp]
\caption{Construction of the High‑Frequency Variable Set $\mathcal{E}_{hf}$}\label{alg:hf_sys}
\KwIn{$\mathcal{C}$: set of constraints; \\
      $freq$: occurrence count of variables; \\
      $\lambda$: frequency threshold; \\
      $\mathcal{E}_{eq}$: equality‑system variables}
\KwOut{$\mathcal{E}_{hf}$: high‑frequency system variables}

$\mathcal{C}_{\neq} \leftarrow \mathcal{C}\setminus equality\_constraints(\mathcal{C})$\;

$\mathcal{E}_{hf} \leftarrow \{\,v \mid freq(v)>\lambda\,\}$\;

\Repeat{no variable is added to $\mathcal{E}_{hf}$}{
    \ForEach{$c \in \mathcal{C}_{\neq}$}{
        \If{$\operatorname{vars}(c)\cap\mathcal{E}_{hf}\neq \emptyset$}{
            $\mathcal{E}_{hf}\leftarrow \mathcal{E}_{hf}\cup\operatorname{vars}(c)$
        }
    }
}

$\mathcal{E}_{hf}\leftarrow \mathcal{E}_{hf}\setminus \mathcal{E}_{eq}$\;

\Return $\mathcal{E}_{hf}$\;
\end{algorithm}

In LS-LIA, each operation modifies the value of variable $x$ to satisfy a literal $\ell$ that contains $x$. However, when $x$ appears in multiple literals, modifying $x$'s value may inadvertently disrupt the satisfaction of some of these literals. 

For the sake of clarity in the following discussion, we define the frequency of a variable $x$, denoted as $freq(x)$, as the number of literals that contain $x$. If $freq(x) > \lambda$, $x$ is considered a high-frequency variable, where $\lambda$ is a hyperparameter. Therefore, during the search process, it is crucial to minimize operations on high-frequency variables whenever possible.

Algorithm~\ref{alg:hf_sys} builds the high-frequency variable set $\mathcal{E}_{hf}$ in three stages. First, it selects every variable that $freq(x) > \lambda$ to form an initial seed (line 2). Next, it repeatedly expands this seed: while scanning the non-equality constraints, if any such constraint shares a variable with the current $\mathcal{E}_{hf}$, all variables in that constraint are merged into $\mathcal{E}_{hf}$; the iteration stops once a complete scan adds no new variables, yielding a transitive closure (lines 3-7). Finally, variables already assigned to the equality system $\mathcal{E}_{eq}$ are removed, ensuring that each shared variable is handled only by the equality subsystem (line 8).

When initializing the variables in the high-frequency system, we calculate their value ranges based on the model obtained from the CDCL(T) phase and then randomly assign values to the variables within those ranges. Specifically, for each variable $x$ to be initialized in the high-frequency system, we define the set of inequalities containing \( x \) that are true under the model \( M \) obtained in the CDCL(T) phase as \(lit_{true}(x) \). For each \( \ell_i \in lit_{true}(x)\), we substitute the values from \( M \) to calculate a feasible interval \( I_i \) for \( x \) under the interpretation of \( M \) with respect to \(\ell_i\). Therefore, each \( \ell_i \) corresponds to a feasible interval \( I_i \). Subsequently, the feasible intervals are merged based on the logical relationships between the literals: for conjunctions of literals \( \ell_i \) and \( \ell_j \), the intervals are intersected, i.e., \( I(x) = I_i \cap I_j \); for disjunctions of literals \( \ell_i \) and \( \ell_j \), the intervals are united, i.e., \( I(x) = I_i \cup I_j \). Finally, we obtain the assignment interval \( I(x) \) for \( x \). A specific example is given in Example~\ref{ex:cdcl_initial}.

\begin{example}\label{ex:cdcl_initial}
Given the SMT(LIA) formula 
\begin{align*}
F = (\ell_1 \lor \ell_2) \land \ell_3 \land \ell_4 = & (x_2 - x_1 \le -1 \lor -x_1 \le -10) \land \\
  & (x_2 - x_3 \le 0) \land (x_1 - x_3 \le 3)
\end{align*}
and the model $M := \{ x_1 = 10, x_2 = 7, x_3 = 7 \}$.
The literals \( \{\ell_1, \ell_2, \ell_4\} \), which contain \( x_1 \) and are true under the model \( M \), are used to compute the initialization interval of \( x_1 \) as follows:

$\ell_1 := x_1 \ge x_2 + 1$, corresponding to $I_1 := [8, +\infty]$.

$\ell_2 := x_1 \ge 10$, corresponding to $I_2 := [10, +\infty]$.

$\ell_4 := x_1 \le x_3 + 3$, corresponding to $I_4 := [-\infty, 10]$.

Therefore, the initialization range for $x_1$ is $I(x_1) := (I_1 \cup I_2) \cap I_4 = [8, 10]$.
\end{example}

It is important to note that in our implementation, the maximum value of a 64-bit signed integer ($2^{63} - 1$) represents positive infinity, and the minimum value ($-2^{63}$) represents negative infinity for unbounded integer variables.

\subsubsection{Initialization of Variables in General System}

Apart from the cases mentioned above, other constraints are generally easier to satisfy, and their initial states have a smaller impact on subsequent searches. Therefore, for the variables in these constraints, their values can be initialized to be far from the model obtained in the previous round of sampling.

Referring to Example~\ref{ex:cdcl_initial}, if \( x_1 \) is a general variable, we initialize it with the complement of the computed result, i.e., \( [-\infty, 7] \cup [11, +\infty] \).

\subsection{Boundary-Aware Move}
Previous research on local search for \SMTLIA has introduced the \emph{critical move} (cm) operator.
Given a literal $\ell$ containing $x$ and currently falsified, each operation $cm(x, \ell)$ modifies the assignment of $x$ to make $\ell$ satisfied~\cite{cai2022local}. However, this operator always modifies the variable $x$ to the boundary value that satisfies $\ell$, which clearly imposes a strong restriction on the possible modified values of the variable.

\begin{figure}[H]
    \centering
    \includegraphics[width=\columnwidth, trim=8cm 9cm 11cm 7.8cm, clip]{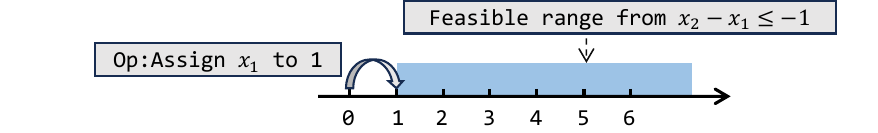}
    \caption{When performing the operation $cm(x_1,x_2-x_1 \le -1)$, any assignment of the variable $x_1$ within $[1, +\infty]$ will make $x_2-x_1 \le -1$ true.}
    \label{fig:critical_move_example}
\end{figure}

The possible modified values of $x$ are not unique to make $\ell$ satisfied. 
For example, consider the formula in Example~\ref{ex:cdcl_initial} with the current assignment $\alpha := \{x_1 = 0, x_2 = 0, x_3 = 0\}$. As illustrated in Figure~\ref{fig:critical_move_example}, performing the operation $cm(x_1, x_2 - x_1 \leq -1)$ assigns $x_1$ to 1. In fact, any value within the interval $[1, +\infty]$ for $x_1$ will satisfy $x_2 - x_1 \leq -1$. However, note that when $x_1$ is assigned a value within the interval $[4, +\infty]$, the constraint $x_1 - x_3 \leq 3$ is no longer satisfied.
We need to reach a good balance between search efficiency and solution diversity in \SMTLIA sampling. 
Thus, we introduce a new operator called \BAM (bam) to replace the cm operator in \HighDiv.

To facilitate describing the new operator, we define \textit{contextual literals} for the integer variable \( x \) as \( \mathcal{L}^{\text{ctx}}(x) = \{ \ell \mid \ell \text{ contains $x$ and is satisfied under $\alpha$} \} \).

\begin{definition}\label{def:bam}
The \BAM operator, denoted as $bam(x, \ell, \mathcal{L}^{\text{ctx}}(x))$, assigns an integer value $c$ to the variable $x$ to satisfy the currently falsified literal $\ell$. The value of $c$ is uniformly sampled within the interval determined by $\ell$ and $\mathcal{L}^{\text{ctx}}(x)$ under the current assignment $\alpha$.
\end{definition}

First, we describe how to calculate the sampling interval determined by $\ell$ and its contextual literals. Each time the operation $bam(x, \ell, \mathcal{L}^{\text{ctx}}(x))$ is performed, we first compute the minimal adjustment $\delta \neq 0$ to the current assignment $\alpha(x)$ needed to satisfy the literal $\ell$. If $\delta > 0$, then $\alpha(x) + \delta$ is a lower bound for the sampling interval; if $\delta < 0$, then $\alpha(x) + \delta$ is the upper bound for the sampling interval, ensuring $\ell$ becomes true after the operation.
Then, for each contextual literal $\ell_i \in \mathcal{L}^{\text{ctx}}(x)$, selected with a certain probability (the probability calculation method will be introduced later), we substitute the values of all variables other than $x$ from $\alpha$ and compute the resulting bound for $x$. Once the lower bound (or upper bound) is established, only the contextual literals $\ell_i$ that tighten the opposite bound need to be considered. By fixing one boundary and focusing on constraining only the opposite boundary, this approach effectively reduces computational overhead while preserving a larger feasible interval for the variable.

When multiple upper bounds (or lower bounds) are derived from contextual literals, the tightest upper bound (i.e., the smallest) or the greatest lower bound (i.e., the largest) should be retained. Moreover, since the primary goal of this move is to satisfy $\ell$, any boundary derived from contextual literals that cannot satisfy $\ell$ should be discarded.

To further balance solution diversity and search efficiency, we introduce a probabilistic mechanism, where the probability $P(\ell_i)$ is used to incorporate the boundary corresponding to the contextual literal $\ell_i$ into the sampling interval calculation. The calculation method for $P(\ell_i)$ is as follows:
\[
    P(\ell) = 1- \frac {step_{\text{sat}}(\ell)} {step_{\text{total}}} \text{,}
\]
where $step_{\text{sat}}(\ell)$ denotes the number of search iterations during which the literal $\ell$ has been satisfied, and $step_{\text{total}}$ is the cumulative number of iterations performed up to the current point.

Intuitively, for literals that are difficult to satisfy, their satisfaction status should be preserved as much as possible after the operation; for literals that are easier to satisfy, attempts can be made to break their constraints.

\begin{figure}[h!]
    \centering
    \includegraphics[width=\columnwidth, trim=8.8cm 9cm 8.8cm 7.8cm, clip]{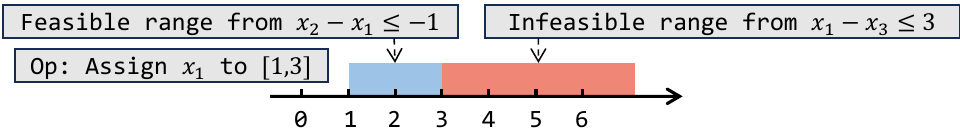}
    \caption{During the operation $bam\!(x,\ell_1,\{\ell_4\})$, the infeasible region specified by $\ell_4 := x_1 - x_3 \le 3$ restricts the admissible range of $x_1$.}
    \label{fig:boundary_aware_move}
\end{figure}

Referring to the formula given in Example~\ref{ex:cdcl_initial}, the current assignment is $\alpha := \{x_1 = 0, x_2 = 0, x_3 = 0\}$. Assume that $P(\ell_4) = 1$. When performing the operation $bam(x_1, \ell_1, \{\ell_4\})$, the initial step involves determining the minimal lower bound to make $\ell_1$ true, resulting in $x_1 \ge 1$. As illustrated in Figure~\ref{fig:boundary_aware_move}, the relevant boundaries are determined from the contextual literal $\ell_4$, giving $x_1 \le 3$. Therefore, $bam(x_1, \ell_1, \{\ell_4\})$ will assign $x_1$ to a random integer value within the interval $[1, 3]$.

\section{Experiments} \label{sec:experiment}
\subsection{Setup}
We will describe the compared state-of-the-art \SMTLIA sampling methods, the benchmark suites, the evaluation metrics, and the research questions. All experiments were conducted on a workstation equipped with a 13th-Generation Intel® Core™ i7-13700F CPU and 32 GB RAM.

\subsubsection{Competitors}
In our experiments, we compare \ensuremath{\mathit{HighDiv}}~\footnote{https://github.com/HighDiv2025/HighDiv2025} with the state-of-the-art \SMTLIA sampling tool, MeGASampler~\footnote{https://github.com/chaosite/MeGASampler}~\cite{peled2023smt}, as well as SMTSampler(Int), the integer logic variant of SMTSampler~\footnote{https://github.com/batchenRothenberg/SMTSampler}~\cite{dutra2018smtsampler}.

\textbf{MeGASampler} is a recently proposed \SMTLIA sampling tool and is also the most advanced tool for \SMTLIA sampling. As noted in its original paper, MeGASampler achieves higher coverage compared to SMTSampler(Int) on most \LIA~benchmarks. Additionally, MeGASampler implements two distinct strategies, both of which we evaluate in our experiments. 

\textbf{SMTSampler(Int)} has been shown to achieve higher coverage than MeGASampler on certain benchmarks, as noted in the paper~\cite{peled2023smt}. Therefore, it will also be evaluated in our experiments.

\subsubsection{Benchmarks}
To verify the effectiveness of the proposed method, we evaluated it on the same \LIA benchmark suite previously used to evaluate MeGASampler~\cite{peled2023smt}. These benchmarks, sourced from the \LIA directory in SMT-LIB~\cite{BarFT-SMTLIB}, were filtered to exclude unreasonable cases based on the following criteria: (1) marked as unsatisfiable or unknown; (2) unable to produce at least 100 samples with any technique; and (3) requiring more than one minute to solve using an SMT solver~\cite{peled2023smt}. Following their methodology, they randomly selected 15 representative benchmarks from each of the 9 directories, resulting in a total of 345 benchmarks to ensure the experiment's tractability, as benchmarks within the same directory tend to be similar.

\subsubsection{Metric}
The diversity of solutions is determined by how comprehensively the sampling results cover the solution space. We have adopted an evaluation metric proposed in prior research, specifically the coverage of internal nodes in Abstract Syntax Trees (ASTs). This method treats a Boolean node as one bit, and for an integer node of any size, only the last 64 bits are considered. For a bit in a variable, if it is sampled as both 1 and 0 in the sample set, we consider that bit to be covered~\cite{peled2023smt}. Coverage is defined as the ratio of covered bits to the total number of bits. For the \SMTLIA sampling problem, this coverage metric reflects, to some extent, how well the sample set explores various logical paths in the program.

\subsubsection{Research Questions}
To assess the effectiveness and efficiency of \HighDiv, we have established the following research questions (RQs). 

\textbf{RQ1:} Can \HighDiv generate a sample set with a higher coverage than its most advanced competitors (i.e., MeGASampler and SMTSampler(Int)) within a fixed time frame?

In this RQ, we adopted the experimental setup from the prior study~\cite{peled2023smt}, with a time limit of 900 seconds, and compared the coverage of sample sets generated by \HighDiv and its competitors.

\textbf{RQ2:} Can \HighDiv achieve a higher coverage than its leading competitors (i.e., MeGASampler and SMTSampler(Int)) in a fixed sample set size?

In this RQ, we compare the performance of \HighDiv with its competitors using a fixed-size sample set. In practice, solutions generated via sampling are typically embedded in the test suite, and executing even a single test campaign is computationally expensive. Therefore, a fixed-size set of test cases is commonly adopted for evaluation~\cite{luo2021ls}.

\textbf{RQ3:} How effective is the core algorithm mechanism of \HighDiv?

This RQ evaluates the effectiveness of $\mathit{HighDiv}$’s core algorithmic components: the \textit{isolation-based variable initialization strategy}, the \BAM operator, and the additional benefits conferred by the stochastic CDCL(T).

\textbf{RQ4:} How do hyperparameter settings influence the performance of \HighDiv?

In this RQ, we analyze how the setting of the hyper-parameter $\lambda$ in the high-frequency system affects the performance of \HighDiv.

\subsection{Experimental Results}
In this section, we first present the results of the experiment and then discuss threats to its validity.

\subsubsection{RQ1: Coverage Comparison within Fixed Time Limits}

To comprehensively compare state-of-the-art \SMTLIA sampling methods, we followed the experimental setup of Peled et al.~\cite{peled2023smt}, setting the time limit to 900 seconds.


\begin{table}[htbp]
	\centering
	\small  
	\begin{tabular}{@{}l cccc @{}}
		\toprule
		Benchmarks & \multicolumn{4}{c}{Coverage (\%)} \\
		\cmidrule(lr){2-5}
		& \HighDiv & MeGA & MeGA\textsuperscript{b} & SMTInt \\
		\midrule
		CAV2009-slacked 	& \textbf{93.15} & 70.15 & 44.77 & 64.24 \\
		CAV2009				& \textbf{76.79} & 44.58 & 69.84 & 55.98 \\
		convert  			& \textbf{20.69}  & 9.23  & 8.48 & 15.02 \\
		dillig        		& \textbf{93.41} & 33.58 & 89.75 & 44.83 \\
		prime-cone       	& \textbf{75.31} & 46.14 & 30.37 & 46.28  \\
		slacks        		& \textbf{95.11} & 71.39 & 47.04 & 64.27  \\
        pb2010 				& \textbf{4.45}  & 4.41  & 4.39 & 2.92   \\
        bofill-sched-random & 11.06 & \textbf{13.14} & 9.56 & 9.86   \\
		bofill-sched-real   & 10.30 & \textbf{11.46} & 9.72 & 8.85   \\
		\midrule
		Average Coverage  & \textbf{53.36} & 33.79 & 34.88 & 34.69 \\
		\bottomrule
	\end{tabular}
	\caption{Comparative Results (Averaged) Across the Benchmarks (Fixed Time 900s).}
    \label{tab:900s_comparison}
\end{table}

Table~\ref{tab:900s_comparison} presents the average coverage comparison among \HighDiv, MeGASampler, and SMTSampler(Int) across 9 selected benchmark folders. To conserve space, only the average values for these 9 folders are shown. \textbf{MeGA} refers to the MeGASampler method based on random sampling, while \textbf{MeGA\textsuperscript{b}} denotes the blocking-based variant. \textbf{SMTInt} indicates the integer logic version of SMTSampler. The best coverage results are highlighted in \textbf{bold}.

As shown in Table~\ref{tab:900s_comparison}, \HighDiv achieves higher coverage than its competitors across most benchmark categories. However, for two \textbf{bofill} categories, \HighDiv achieves lower coverage than MeGA. We attribute this result to the high proportion of equality constraints in these instances, which limits \ensuremath{\mathit{HighDiv}}'s ability to introduce diversity during the search process. However, constraints collected from real-world programs are typically not as stringent.

\begin{figure*}[!t]
	\centering
	\begin{subfigure}[b]{0.32\textwidth}
		\includegraphics[width=\textwidth]{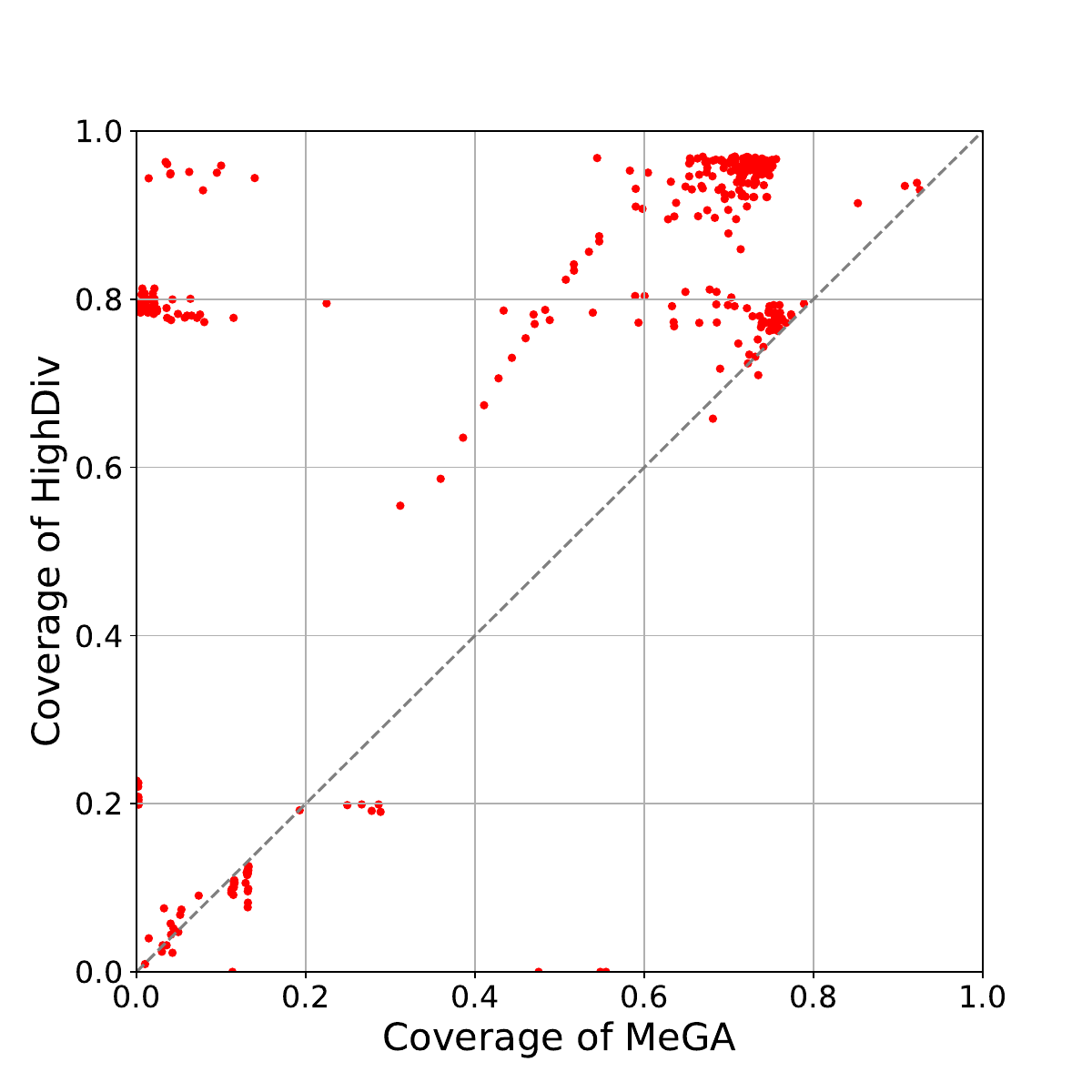}
		\caption{\HighDiv vs MeGA}
		\label{fig:time_limit_sub2_1}
	\end{subfigure}
	\begin{subfigure}[b]{0.32\textwidth}
		\includegraphics[width=\textwidth]{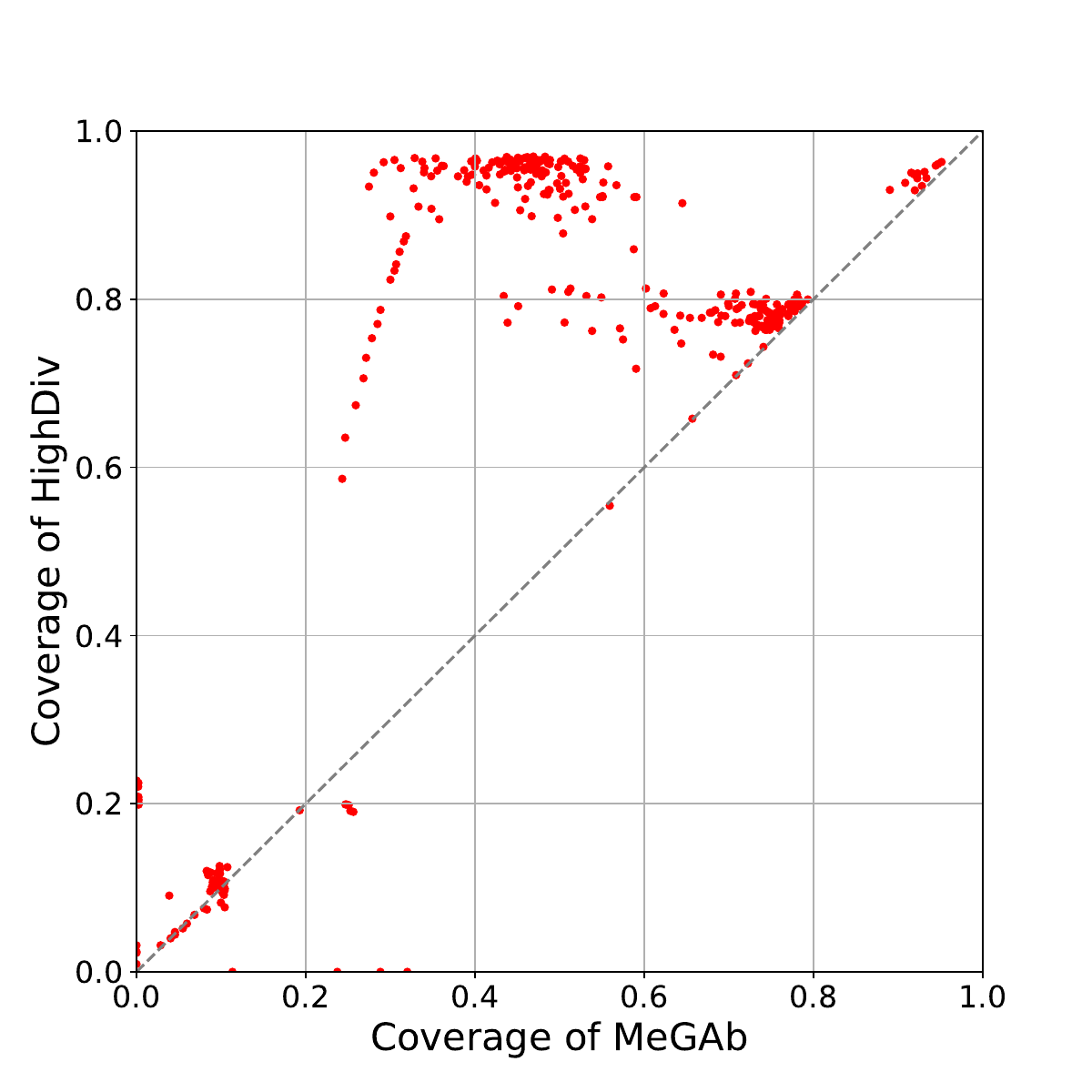}
		\caption{\HighDiv vs MeGA\textsuperscript{b}}
		\label{fig:time_limit_sub2_2}
	\end{subfigure}
	\begin{subfigure}[b]{0.32\textwidth}
		\includegraphics[width=\textwidth]{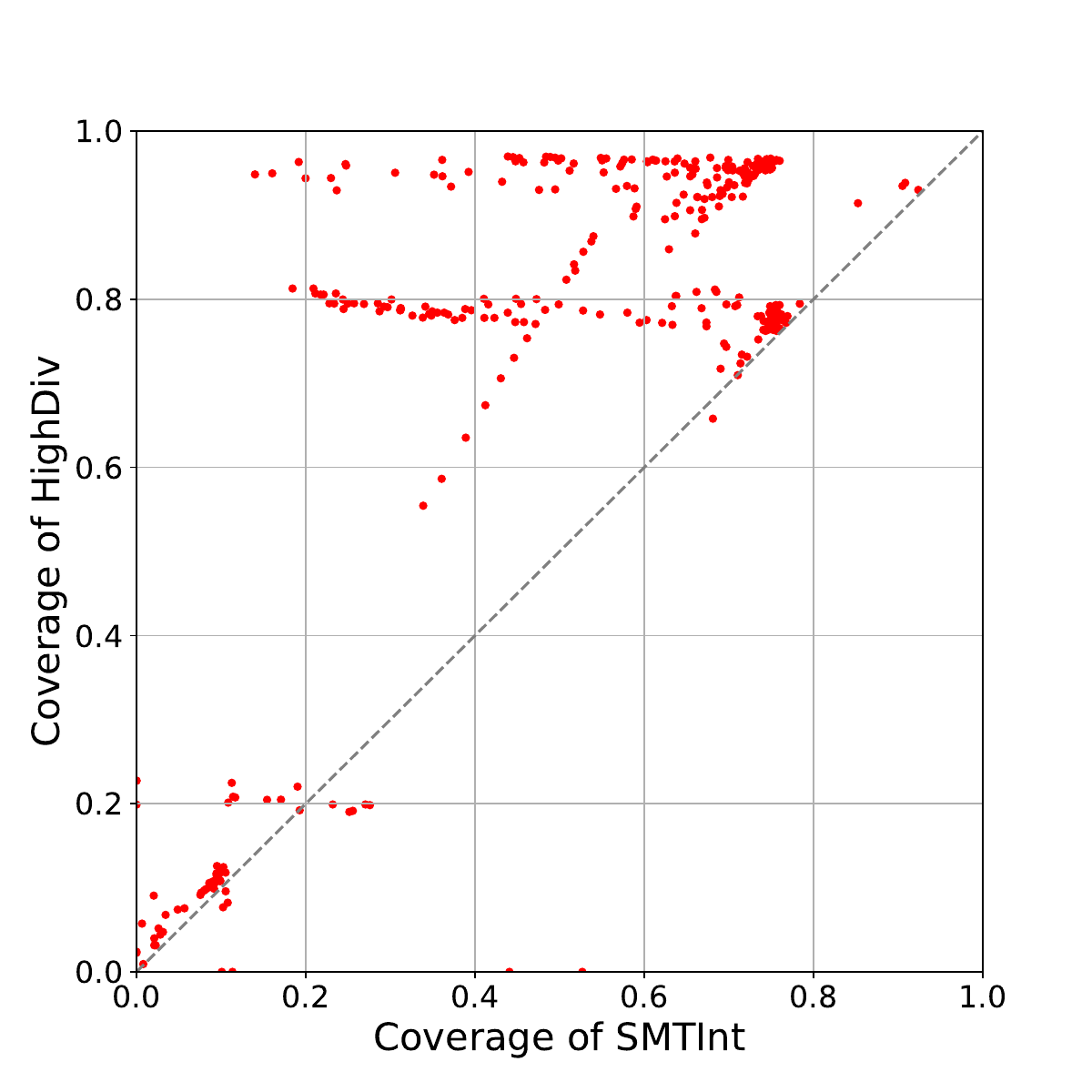}
		\caption{HighDiv vs SMTInt}
		\label{fig:time_limit_sub2_3}
	\end{subfigure}
	\caption{Comparative Coverage Performance of \HighDiv Against Competitors (t = 900 seconds).}
	\label{fig:time_limit_900}
\end{figure*}

To present detailed results for each individual benchmark file, we employ scatter plots (see Figure~\ref{fig:time_limit_900}). It can be observed that, under the same time limit, \HighDiv yields higher coverage in most benchmarks compared to MeGASampler and SMTSampler(Int).

\begin{figure*}[!htbp]
	\centering
	\begin{subfigure}[b]{0.32\textwidth}
		\includegraphics[width=\textwidth]{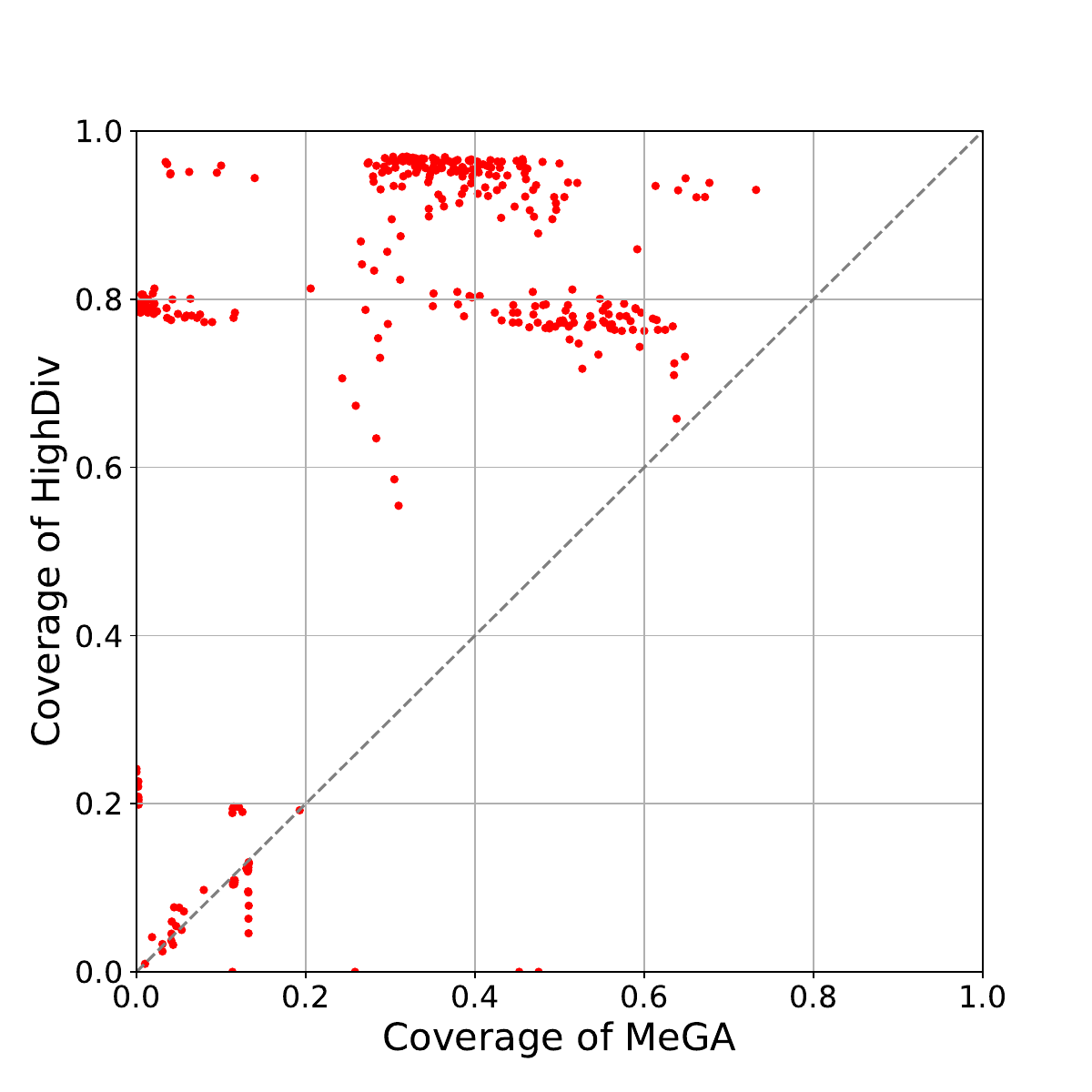}
		\caption{\HighDiv vs MeGA}
		\label{fig:num_limit_sub2_1}
	\end{subfigure}
	\begin{subfigure}[b]{0.32\textwidth}
		\includegraphics[width=\textwidth]{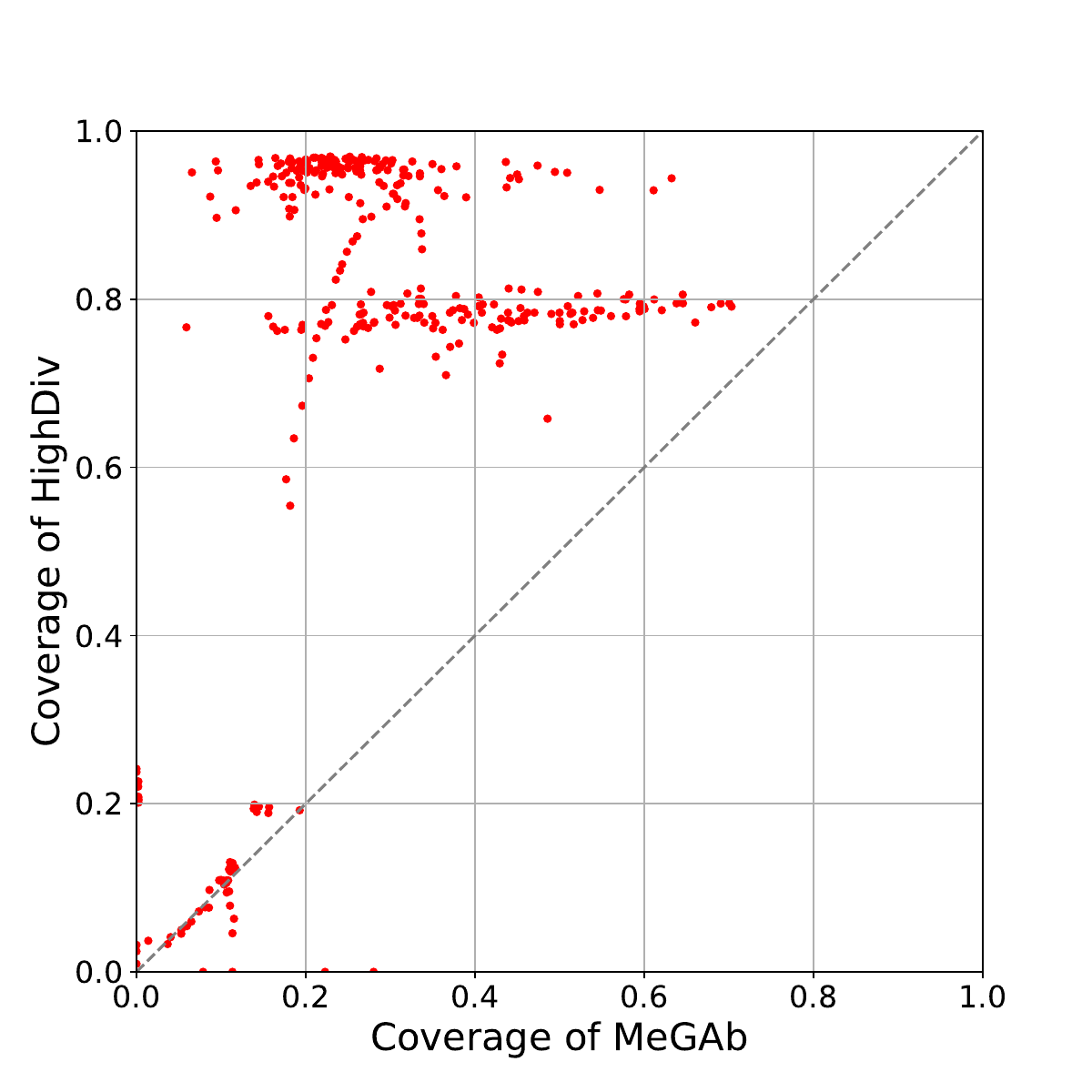}
		\caption{\HighDiv vs MeGA\textsuperscript{b}}
		\label{fig:num_limit_sub2_2}
	\end{subfigure}
	\begin{subfigure}[b]{0.32\textwidth}
		\includegraphics[width=\textwidth]{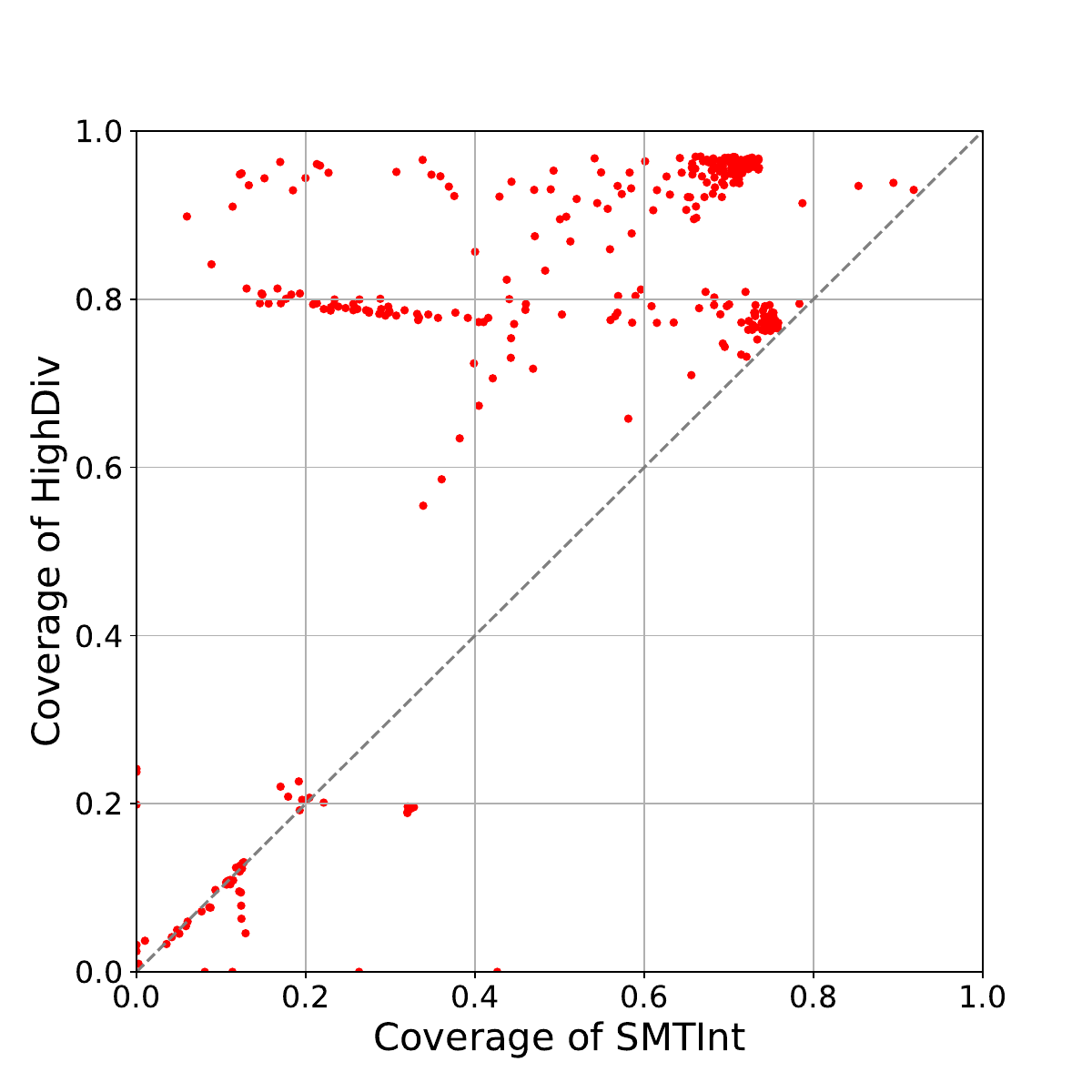}
		\caption{\HighDiv vs SMTInt}
		\label{fig:num_limit_sub2_3}
	\end{subfigure}
	\caption{Comparative Coverage Performance of \HighDiv Against Competitors (k = 1000).}
	\label{fig:num_limit_1000}
\end{figure*}

\subsubsection{RQ2: Comparison of Coverage for Fixed-Size Solution Sets}


\begin{table}[!t]
	\centering
	\small  
	\begin{tabular}{@{}l cccc@{}}
		\toprule
		Benchmarks & \multicolumn{4}{c}{Coverage (\%)} \\
		\cmidrule(lr){2-5} 
		& \HighDiv & MeGA & MeGA\textsuperscript{b} & SMTInt \\
		\midrule
		CAV2009-slacked     & \textbf{93.16} & 38.03 & 23.56 & 63.71 \\
		CAV2009             & \textbf{76.79} & 34.22 & 40.74 & 52.04 \\
		convert             & \textbf{20.78} & 4.04  & 5.95  & 19.84  \\
		dillig              & \textbf{93.41} & 32.87 & 42.19 & 41.01  \\
		prime-cone          & \textbf{75.30} & 28.49 & 21.94 & 40.60  \\
		slacks              & \textbf{95.11} & 39.73 & 25.41 & 62.64  \\
            pb2010              & 4.73  & 4.68   & 5.70  & \textbf{5.09}   \\
            bofill-sched-random & 10.84 & \textbf{13.20} & 11.16 & 12.35 \\
		bofill-sched-real   & 10.72 & \textbf{11.52} & 10.38 & 11.00 \\
            \midrule
		Average Coverage  & \textbf{53.43} & 22.98 & 20.78 & 34.25 \\
		\bottomrule
	\end{tabular}
	\caption{Comparative Results (Averaged) Across the Benchmarks (Fixed Size 1000).}
        \label{tab:1000num_comparison}
\end{table}

Following prior SAT sampling studies, we set the solution set size to $k = 1000$~\cite{luo2021ls} and limited the sampling time to one hour. It should be noted that, due to the stringent constraints of some benchmarks, it was not possible to generate the target number of valid solutions within one hour, so we included all solutions from these benchmarks in our analysis.

From Table~\ref{tab:1000num_comparison}, it can be seen that with $k=1000$, \HighDiv significantly outperforms MeGASampler and SMTSampler(Int) in terms of coverage in most benchmarks. For each benchmark file within the benchmark folders, we still present a scatter plot (see Figure~\ref{fig:num_limit_1000}) comparing the coverage of \HighDiv with its competitors.

\subsubsection{RQ3: Effects of Algorithmic Components}
To analyze the effectiveness of each algorithmic component in \HighDiv, we modify \HighDiv to obtain seven alternative versions as follows.

\begin{itemize}
    \item \textbf{init2zero}: In the local search phase, all integer variables are initially set to zero.
    \item \textbf{init2CD}: In the local search phase, all integer variables are initialized through guidance from CDCL(T).
    \item \textbf{init2RCD}: In the local search phase, all integer variables are initialized through reverse guidance from CDCL(T).
    \item \textbf{init2R}: In the local search phase, all integer variables are initialized randomly.
    \item \textbf{no\_prob}: Removing the probabilistic mechanism in the \BAM operator.
    \item \textbf{no\_bam}: Removing the \BAM operator and using the critical move operator.
    \item \textbf{no\_cdcl}: Removing the stochastic CDCL(T) component and use only the local search component for iteration.
\end{itemize}


\begin{table*}[htbp]
    \centering
    \small
    \begin{tabularx}{\textwidth}{@{}l *{8}{>{\centering\arraybackslash}X}@{}}
        \toprule
        & \HighDiv & init2zero & init2CD & init2RCD & init2R & no\_prob & no\_bam & no\_cdcl \\
        \midrule
        Cov(t = 900sec) & \textbf{53.36\%} & 47.75\% & 38.93\% & 51.09\% & 51.36\% & 49.69\% & 33.82\% & 49.27\% \\
        Cov(k = 1000)   & \textbf{53.43\%} & 47.42\% & 37.90\% & 51.15\% & 51.51\% & 49.47\% & 33.44\% & 50.30\% \\
        \bottomrule
    \end{tabularx}
    \caption{Ablation study in showing the effectiveness of each algorithmic component.}
    \label{tab:ablation}
\end{table*}

Table~\ref{tab:ablation} presents the average coverage of each version across 9 benchmark categories. It can be observed that, under both the 900-second time limit and the 1000-sample constraint, \HighDiv achieves higher coverage compared to the various variants. This demonstrates the effectiveness of each core algorithmic component in \HighDiv.

\subsubsection{RQ4: Effect of Hyper-Parameter Setting}

Table~\ref{tab:hyper} presents the average coverage achieved by \HighDiv across all benchmarks, using different $\lambda$ hyperparameter settings under fixed time and sample size constraints. The results show that, with our default hyperparameter setting ($\lambda = 50$), \HighDiv achieves the highest coverage in both experimental setups. This aligns with our intuition: when $\lambda$ is small, the initial assignment in the local search is closer to the solution from the previous iteration, which enables faster solving but sacrifices solution diversity. In contrast, when $\lambda$ is large, the initial assignment is more distant from the previous model, enhancing solution diversity but negatively impacting solving efficiency.

\subsection{Threats to Validity}
The validity of our evaluation may be subject to the following potential threats:

\textbf{Correctness of Implementation.}

To ensure the correctness of our implementation, we validated all generated samples to confirm their satisfaction of the given formulas, thereby guaranteeing the robustness and accuracy of the results.

\textbf{Validity of the Experimental Setup.}

Different experimental configurations may pose potential threats to the validity of our evaluation. Therefore, following previous studies~\cite{peled2023smt}, we conducted experiments with a time limit of 900 seconds; and, in accordance with recent recommendations~\cite{luo2021ls}, we set the sample size $k$ to 1000 to mitigate this risk. The results in Tables 1 and 2 demonstrate that under both configurations, \HighDiv significantly outperforms MeGASampler and SMTSampler(Int) in terms of diversity on most benchmarks.

\textbf{Generality of Benchmarking Instances.}

To mitigate potential threats to the validity of the experiments, we selected nine benchmark instances that have been used in recent studies~\cite{peled2023smt}. These instances are sourced from SMT-LIB~\cite{BarFT-SMTLIB} and have been thoroughly analyzed in numerous related works~\cite{peled2023smt,cai2022local,DBLP:conf/sat/HyvarinenMAS16,ijcai2024p211,DBLP:conf/cav/BjornerN24}. The benchmarks include both relaxed and complex constraints, thereby possessing strong representativeness and broad applicability, effectively reducing potential threats.

\begin{table}[htbp]
    \centering
    \small  
    \begin{tabularx}{\linewidth}{@{}l *{3}{>{\centering\arraybackslash}X}@{}}
        \toprule
        & $\lambda = 20$ & $\lambda = 50$ & $\lambda = 80$ \\
        \midrule
        Cov(t = 900sec) & 48.31\% & \textbf{53.36\%} & 53.27\% \\
        Cov(k = 1000)   & 48.38\% & \textbf{53.43\%} & 53.08\% \\
        \bottomrule
    \end{tabularx}
    \caption{The average coverage of sampling results by \HighDiv with different hyperparameter settings of $\lambda$ across all benchmarks.}
    \label{tab:hyper}
\end{table}

\section{Related Works} \label{sec:related}
The constraint sampling problem is a significant research topic in software testing~\cite{holler2012fuzzing, godefroid2005dart, sen2005cute}. Over the past decades, satisfiability (SAT) sampling has been extensively studied. Current SAT sampling methods include Markov-Chain Monte-Carlo (MCMC)~\cite{kitchen2007stimulus, kitchen2010markov}, universal hashing~\cite{meel2014sampling, meel2016constrained, ermon2013embed}, heuristic local search~\cite{luo2021ls}, and knowledge compilation~\cite{lai2017new}, all of which have effectively addressed SAT sampling challenges. While SMT formula sampling can be tackled using existing SAT methods by encoding SMT formulas into SAT~\cite{shaw2024csb}, this transformation loses the formulas' high-level structure, which could otherwise improve sampling efficiency and diversity~\cite{dutra2019guidedsampler, peled2023smt, dutra2018smtsampler}.

Dutra et al. introduced SMTSampler, the first sampler designed specifically for sampling SMT bit-vector theory formulas~\cite{dutra2018smtsampler}. SMTSampler uses MaxSAT modulo theories (MAX-SMT) to generate initial random seeds and then obtains a set of solutions through syntactic mutation and combination. Subsequent research, GuidedSampler, allows for the setting of coverage metrics specific to certain problems and aims to optimize these metrics~\cite{dutra2019guidedsampler}. These methods primarily handle fixed-width bit-vector theory formulas, but this approach is unable to efficiently sample the linear integer constraints. In response, Peled et al. proposed the first sampling method for \SMTLIA formula, named MeGASampler~\cite{peled2023smt}. This method first uses an existing SMT solver to solve constraints, then generates an under-approximation formula of the original formula by adding extra constraints to the obtained solutions, and finally performs sampling based on this under-approximation formula. This is considered the state-of-the-art solution to \SMTLIA sampling problems. However, these SMT samplers regard SMT solvers as black boxes and fail to introduce diversity into the solving process, thereby limiting the diversity of the generated solutions. Furthermore, these samplers frequently invoke MaxSMT to increase the randomness of the initial seeds, but using MaxSMT is very expensive, thus creating substantial overhead.

In contrast, \HighDiv aims to improve sampling diversity. It presents the first \SMTLIA sampling framework that combines local search with bidirectionally guided CDCL(T) iterations. During the local search phase, we systematically analyze the impact of variable initialization on both sampling diversity and efficiency, achieving an effective balance through an \textit{isolation-based variable initialization strategy}. We also introduce the \BAM operator, which balances diversity and search efficiency throughout the search. As shown in Section~\ref{sec:experiment}, extensive experiments on SMT-LIB standard \LIA benchmarks, including real-world software engineering instances, demonstrate that \HighDiv produces more diverse sample sets than MeGASampler and SMTSampler(Int).
\section{Conclusion} \label{sec:conclusion}
Previous SMT sampling algorithms often struggle to generate diverse samples as they treat the SMT solver as a black box and rely on extending a single model, limiting the coverage of the solution space. To overcome this, we propose \HighDiv, the first sampling framework that combines local search with bidirectional CDCL(T) iterations. In the local search phase, \HighDiv introduces an isolation-based variable initialization strategy and the \BAM operator, effectively balancing sample diversity and search efficiency. Experimental results show that \HighDiv produces significantly more diverse samples than MeGASampler and SMTSampler(Int) within the same time limit.

It is important to note that \HighDiv is orthogonal to existing SMT sampling algorithms such as MeGASampler and SMTSampler. By generating initial models with \HighDiv and then applying model-guided approximation or combinatorial mutation techniques, overall sampling efficiency can be greatly improved. In future work, we plan to extend \HighDiv to support additional theories, such as Linear Real Arithmetic and Quantifier-Free Bit-Vectors, and explore integrating model-guided approximation and combinatorial mutation strategies to further enhance its scalability.

\bibliography{bibliography}

\begin{thebibliography}{10}
\providecommand{\url}[1]{#1}
\csname url@samestyle\endcsname
\providecommand{\newblock}{\relax}
\providecommand{\bibinfo}[2]{#2}
\providecommand{\BIBentrySTDinterwordspacing}{\spaceskip=0pt\relax}
\providecommand{\BIBentryALTinterwordstretchfactor}{4}
\providecommand{\BIBentryALTinterwordspacing}{\spaceskip=\fontdimen2\font plus
\BIBentryALTinterwordstretchfactor\fontdimen3\font minus \fontdimen4\font\relax}
\providecommand{\BIBforeignlanguage}[2]{{%
\expandafter\ifx\csname l@#1\endcsname\relax
\typeout{** WARNING: IEEEtran.bst: No hyphenation pattern has been}%
\typeout{** loaded for the language `#1'. Using the pattern for}%
\typeout{** the default language instead.}%
\else
\language=\csname l@#1\endcsname
\fi
#2}}
\providecommand{\BIBdecl}{\relax}
\BIBdecl

\bibitem{peleska2011automated}
J.~Peleska, E.~Vorobev, and F.~Lapschies, ``Automated test case generation with smt-solving and abstract interpretation,'' in \emph{{NASA} Formal Methods - Third International Symposium, {NFM} 2011}, ser. Lecture Notes in Computer Science, M.~G. Bobaru, K.~Havelund, G.~J. Holzmann, and R.~Joshi, Eds., vol. 6617, 2011, pp. 298--312.

\bibitem{cadar2008klee}
C.~Cadar, D.~Dunbar, and D.~R. Engler, ``{KLEE:} unassisted and automatic generation of high-coverage tests for complex systems programs,'' in \emph{8th {USENIX} Symposium on Operating Systems Design and Implementation, {OSDI} 2008}, R.~Draves and R.~van Renesse, Eds., 2008, pp. 209--224.

\bibitem{DBLP:conf/uss/PoeplauF20}
S.~Poeplau and A.~Francillon, ``Symbolic execution with symcc: Don't interpret, compile!'' in \emph{29th {USENIX} Security Symposium, {USENIX} Security 2020}, S.~Capkun and F.~Roesner, Eds., 2020, pp. 181--198.

\bibitem{godefroid2008automated}
P.~Godefroid, M.~Y. Levin, and D.~A. Molnar, ``Automated whitebox fuzz testing,'' in \emph{Proceedings of the Network and Distributed System Security Symposium, {NDSS} 2008}.\hskip 1em plus 0.5em minus 0.4em\relax The Internet Society, 2008.

\bibitem{godefroid2005dart}
P.~Godefroid, N.~Klarlund, and K.~Sen, ``{DART:} directed automated random testing,'' in \emph{Proceedings of the {ACM} {SIGPLAN} 2005 Conference on Programming Language Design and Implementation}, V.~Sarkar and M.~W. Hall, Eds., 2005, pp. 213--223.

\bibitem{jiang2023evaluating}
L.~Jiang, H.~Yuan, M.~Wu, L.~Zhang, and Y.~Zhang, ``Evaluating and improving hybrid fuzzing,'' in \emph{45th {IEEE/ACM} International Conference on Software Engineering, {ICSE} 2023}, 2023, pp. 410--422.

\bibitem{liu2020legion}
D.~Liu, G.~Ernst, T.~Murray, and B.~I.~P. Rubinstein, ``{LEGION:} best-first concolic testing,'' in \emph{35th {IEEE/ACM} International Conference on Automated Software Engineering, {ASE} 2020}, 2020, pp. 54--65.

\bibitem{huang2020pangolin}
H.~Huang, P.~Yao, R.~Wu, Q.~Shi, and C.~Zhang, ``Pangolin: Incremental hybrid fuzzing with polyhedral path abstraction,'' in \emph{2020 {IEEE} Symposium on Security and Privacy, {SP} 2020}, 2020, pp. 1613--1627.

\bibitem{dutra2018smtsampler}
R.~Dutra, J.~Bachrach, and K.~Sen, ``Smtsampler: efficient stimulus generation from complex {SMT} constraints,'' in \emph{Proceedings of the International Conference on Computer-Aided Design, {ICCAD} 2018}, I.~Bahar, Ed., 2018, p.~30.

\bibitem{dutra2019guidedsampler}
R.~Dutra, J.~Bachrach, and K.~Sen, ``{GUIDEDSAMPLER:} coverage-guided sampling of {SMT} solutions,'' in \emph{2019 Formal Methods in Computer Aided Design, {FMCAD} 2019}, C.~W. Barrett and J.~Yang, Eds., 2019, pp. 203--211.

\bibitem{peled2023smt}
M.~Peled, B.~Rothenberg, and S.~Itzhaky, ``{SMT} sampling via model-guided approximation,'' in \emph{Formal Methods - 25th International Symposium, {FM} 2023}, ser. Lecture Notes in Computer Science, M.~Chechik, J.~Katoen, and M.~Leucker, Eds., vol. 14000, 2023, pp. 74--91.

\bibitem{jfs-sampler:icst25}
M.~Carrasco, C.~Cadar, and A.~Donaldson, ``Scalable smt sampling for floating-point formulas via coverage-guided fuzzing,'' in \emph{IEEE International Conference on Software Testing, Verification, and Validation (ICST 2025)}, 2025.

\bibitem{DBLP:conf/qrs/RobertGWS21}
C.~Robert, J.~Guiochet, H.~Waeselynck, and L.~V. Sartori, ``{TAF:} a tool for diverse and constrained test case generation,'' in \emph{21st {IEEE} International Conference on Software Quality, Reliability and Security, {QRS} 2021}, 2021, pp. 311--321.

\bibitem{DBLP:journals/ese/HeradioFGBB22}
R.~Heradio, D.~Fern{\'{a}}ndez{-}Amor{\'{o}}s, J.~A. Galindo, D.~Benavides, and D.~S. Batory, ``Uniform and scalable sampling of highly configurable systems,'' \emph{Empir. Softw. Eng.}, vol.~27, no.~2, p.~44, 2022.

\bibitem{mccarthy1993towards}
J.~McCarthy, ``Towards a mathematical science of computation,'' in \emph{Information Processing, Proceedings of the 2nd {IFIP} Congress 1962}, 1962, pp. 21--28.

\bibitem{de2008z3}
L.~M. de~Moura and N.~S. Bj{\o}rner, ``{Z3:} an efficient {SMT} solver,'' in \emph{Tools and Algorithms for the Construction and Analysis of Systems, 14th International Conference, {TACAS} 2008}, ser. Lecture Notes in Computer Science, C.~R. Ramakrishnan and J.~Rehof, Eds., vol. 4963, 2008, pp. 337--340.

\bibitem{barbosa2022cvc5}
H.~Barbosa, C.~W. Barrett, M.~Brain, G.~Kremer, H.~Lachnitt, M.~Mann, A.~Mohamed, M.~Mohamed, A.~Niemetz, A.~N{\"{o}}tzli, A.~Ozdemir, M.~Preiner, A.~Reynolds, Y.~Sheng, C.~Tinelli, and Y.~Zohar, ``cvc5: {A} versatile and industrial-strength {SMT} solver,'' in \emph{Tools and Algorithms for the Construction and Analysis of Systems - 28th International Conference, {TACAS} 2022}, ser. Lecture Notes in Computer Science, D.~Fisman and G.~Rosu, Eds., vol. 13243, 2022, pp. 415--442.

\bibitem{cimatti2013mathsat5}
A.~Cimatti, A.~Griggio, B.~J. Schaafsma, and R.~Sebastiani, ``The mathsat5 {SMT} solver,'' in \emph{Tools and Algorithms for the Construction and Analysis of Systems - 19th International Conference, {TACAS} 2013}, ser. Lecture Notes in Computer Science, N.~Piterman and S.~A. Smolka, Eds., vol. 7795, 2013, pp. 93--107.

\bibitem{luo2021ls}
C.~Luo, B.~Sun, B.~Qiao, J.~Chen, H.~Zhang, J.~Lin, Q.~Lin, and D.~Zhang, ``Ls-sampling: an effective local search based sampling approach for achieving high t-wise coverage,'' in \emph{{ESEC/FSE} '21: 29th {ACM} Joint European Software Engineering Conference and Symposium on the Foundations of Software Engineering}, D.~Spinellis, G.~Gousios, M.~Chechik, and M.~D. Penta, Eds., 2021, pp. 1081--1092.

\bibitem{luo2024solving}
C.~Luo, J.~Song, Q.~Zhao, B.~Sun, J.~Chen, H.~Zhang, J.~Lin, and C.~Hu, ``Solving the \emph{t}-wise coverage maximum problem via effective and efficient local search-based sampling,'' \emph{{ACM} Trans. Softw. Eng. Methodol.}, vol.~34, no.~1, pp. 13:1--13:64, 2025.

\bibitem{cai2022local}
S.~Cai, B.~Li, and X.~Zhang, ``Local search for {SMT} on linear integer arithmetic,'' in \emph{Computer Aided Verification - 34th International Conference, {CAV} 2022}, ser. Lecture Notes in Computer Science, S.~Shoham and Y.~Vizel, Eds., vol. 13372, 2022, pp. 227--248.

\bibitem{zhang2024deep}
X.~Zhang, B.~Li, and S.~Cai, ``Deep combination of {CDCL(T)} and local search for satisfiability modulo non-linear integer arithmetic theory,'' in \emph{Proceedings of the 46th {IEEE/ACM} International Conference on Software Engineering, {ICSE} 2024}, 2024, pp. 125:1--125:13.

\bibitem{kroening2016decision}
D.~Kroening and O.~Strichman, \emph{Decision Procedures - An Algorithmic Point of View, Second Edition}, ser. Texts in Theoretical Computer Science. An {EATCS} Series.\hskip 1em plus 0.5em minus 0.4em\relax Springer, 2016.

\bibitem{barrett2021satisfiability}
C.~W. Barrett, R.~Sebastiani, S.~A. Seshia, and C.~Tinelli, ``Satisfiability modulo theories,'' in \emph{Handbook of Satisfiability - Second Edition}, ser. Frontiers in Artificial Intelligence and Applications, A.~Biere, M.~Heule, H.~van Maaren, and T.~Walsh, Eds., 2021, vol. 336, pp. 1267--1329.

\bibitem{ganzinger2004dpll}
H.~Ganzinger, G.~Hagen, R.~Nieuwenhuis, A.~Oliveras, and C.~Tinelli, ``{DPLL(} {T):} fast decision procedures,'' in \emph{Computer Aided Verification, 16th International Conference, {CAV} 2004}, ser. Lecture Notes in Computer Science, R.~Alur and D.~A. Peled, Eds., vol. 3114, 2004, pp. 175--188.

\bibitem{golia2021designing}
P.~Golia, M.~Soos, S.~Chakraborty, and K.~S. Meel, ``Designing samplers is easy: The boon of testers,'' in \emph{Formal Methods in Computer Aided Design, {FMCAD} 2021}, 2021, pp. 222--230.

\bibitem{frohlich2015stochastic}
A.~Fr{\"{o}}hlich, A.~Biere, C.~M. Wintersteiger, and Y.~Hamadi, ``Stochastic local search for satisfiability modulo theories,'' in \emph{Proceedings of the Twenty-Ninth {AAAI} Conference on Artificial Intelligence}, B.~Bonet and S.~Koenig, Eds., 2015, pp. 1136--1143.

\bibitem{liu2025optimizing}
C.~Liu, G.~Liu, C.~Luo, S.~Cai, Z.~Lei, W.~Zhang, Y.~Chu, and G.~Zhang, ``Optimizing local search-based partial maxsat solving via initial assignment prediction,'' \emph{Sci. China Inf. Sci.}, vol.~68, no.~2, 2025.

\bibitem{BarFT-SMTLIB}
C.~Barrett, P.~Fontaine, and C.~Tinelli, ``{The Satisfiability Modulo Theories Library (SMT-LIB)},'' {\tt www.SMT-LIB.org}, 2016.

\bibitem{DBLP:conf/sat/HyvarinenMAS16}
A.~E.~J. Hyv{\"{a}}rinen, M.~Marescotti, L.~Alt, and N.~Sharygina, ``Opensmt2: An {SMT} solver for multi-core and cloud computing,'' in \emph{Theory and Applications of Satisfiability Testing - {SAT} 2016}, ser. Lecture Notes in Computer Science, N.~Creignou and D.~L. Berre, Eds., vol. 9710, 2016, pp. 547--553.

\bibitem{ijcai2024p211}
Z.~Lu, S.~Siemer, P.~Jha, J.~Day, F.~Manea, and V.~Ganesh, ``Layered and staged monte carlo tree search for smt strategy synthesis,'' in \emph{Proceedings of the Thirty-Third International Joint Conference on Artificial Intelligence, {IJCAI-24}}, K.~Larson, Ed., 2024, pp. 1907--1915.

\bibitem{DBLP:conf/cav/BjornerN24}
N.~S. Bj{\o}rner and L.~Nachmanson, ``Arithmetic solving in {Z3},'' in \emph{Computer Aided Verification - 36th International Conference, {CAV} 2024}, ser. Lecture Notes in Computer Science, A.~Gurfinkel and V.~Ganesh, Eds., vol. 14681, 2024, pp. 26--41.

\bibitem{holler2012fuzzing}
C.~Holler, K.~Herzig, and A.~Zeller, ``Fuzzing with code fragments,'' in \emph{Proceedings of the 21th {USENIX} Security Symposium}, T.~Kohno, Ed., 2012, pp. 445--458.

\bibitem{sen2005cute}
K.~Sen, D.~Marinov, and G.~Agha, ``{CUTE:} a concolic unit testing engine for {C},'' in \emph{Proceedings of the 10th European Software Engineering Conference held jointly with 13th {ACM} {SIGSOFT} International Symposium on Foundations of Software Engineering, 2005}, M.~Wermelinger and H.~C. Gall, Eds., 2005, pp. 263--272.

\bibitem{kitchen2007stimulus}
N.~Kitchen and A.~Kuehlmann, ``Stimulus generation for constrained random simulation,'' in \emph{2007 International Conference on Computer-Aided Design, {ICCAD} 2007}, G.~G.~E. Gielen, Ed., 2007, pp. 258--265.

\bibitem{kitchen2010markov}
N.~Kitchen, ``Markov chain monte carlo stimulus generation for constrained random simulation,'' Ph.D. dissertation, University of California, Berkeley, {USA}, 2010.

\bibitem{meel2014sampling}
K.~S. Meel, ``Sampling techniques for boolean satisfiability,'' \emph{CoRR}, vol. abs/1404.6682, 2014.

\bibitem{meel2016constrained}
K.~S. Meel, M.~Y. Vardi, S.~Chakraborty, D.~J. Fremont, S.~A. Seshia, D.~Fried, A.~Ivrii, and S.~Malik, ``Constrained sampling and counting: Universal hashing meets {SAT} solving,'' in \emph{Beyond NP, Papers from the 2016 {AAAI} Workshop}, ser. {AAAI} Technical Report, A.~Darwiche, Ed., vol. {WS-16-05}, 2016.

\bibitem{ermon2013embed}
S.~Ermon, C.~P. Gomes, A.~Sabharwal, and B.~Selman, ``Embed and project: Discrete sampling with universal hashing,'' in \emph{Advances in Neural Information Processing Systems 26: 27th Annual Conference on Neural Information Processing Systems 2013.}, C.~J.~C. Burges, L.~Bottou, Z.~Ghahramani, and K.~Q. Weinberger, Eds., 2013, pp. 2085--2093.

\bibitem{lai2017new}
Y.~Lai, D.~Liu, and M.~Yin, ``New canonical representations by augmenting obdds with conjunctive decomposition (extended abstract),'' in \emph{Proceedings of the Twenty-Sixth International Joint Conference on Artificial Intelligence, {IJCAI} 2017}, C.~Sierra, Ed., 2017, pp. 5010--5014.

\bibitem{shaw2024csb}
A.~Shaw and K.~S. Meel, ``{CSB:} {A} counting and sampling tool for bit-vectors,'' in \emph{Proceedings of the 22nd International Workshop on Satisfiability Modulo Theories co-located with the 36th International Conference on Computer Aided Verification {(CAV} 2024)}, ser. {CEUR} Workshop Proceedings, G.~Reger and Y.~Zohar, Eds., vol. 3725, 2024, pp. 36--43.

\end{thebibliography}

\end{document}